\documentclass[aps,pra,reprint,superscriptaddress]{revtex4-1}

\usepackage{graphicx} 
\usepackage{amsmath,bm}

\begin{document}
% Title of paper
\title{Localization in the constrained quantum annealing of graph coloring} 

% Authors
\author{Kazue Kudo}
\email[]{kudo@is.ocha.ac.jp}
\affiliation{Department of Computer Science, Ochanomizu University, Tokyo 112-8610, Japan}

% Abstract
\begin{abstract}
 Constrained quantum annealing (CQA) is a quantum annealing approach that is designed so that constraints are satisfied without penalty terms. 
There is an analogy between the model for the CQA of graph coloring and a set of disordered spin chains.
In the model for the CQA of graph coloring, disorder corresponds to the fluctuation of effective local fields that increase in a CQA process.
Numerical simulations of effective fields and entanglement demonstrate how localization appears in the CQA.
Some notable features appear in the concurrence, which is a measure of entanglement, plotted as a function of the fluctuation of effective fields.
\end{abstract}

\maketitle

\section{Introduction}
\label{sec:intro}

% Quantum annealing and constrained quantum annealing
Quantum annealing (QA), which is essentially the same as adiabatic quantum computation, has attracted huge interest in recent years.
QA is typically known as a quantum-mechanical approach for combinatorial optimization problems~\cite{Kadowaki1998,Brooke1999,Das2008,Farhi2001,Martonak2004,Kurihara2009,Altshuler2010,Titiloye2011,Titiloye2012}.
In a QA framework, the total Hamiltonian consists of a problem Hamiltonian and a driver Hamiltonian.
The problem Hamiltonian describes an optimization problem and is to be minimized.
The driver Hamiltonian corresponds to quantum fluctuation.
The QA process starts with the ground state of the driver Hamiltonian.
The proportion of the driver Hamiltonian in the total one decreases as time proceeds.
Instead, that of the problem Hamiltonian increases, 
and the total Hamiltonian coincides with the problem Hamiltonian in the end.
If the process is adiabatic, the solution to the problem, 
i.e., the ground state of the problem Hamiltonian, 
is obtained as the ground state of the total Hamiltonian at the end of the QA process.
The computational efficiency of QA depends on the energy gap between the ground and the first-excited states.
Several approaches have been proposed to avoid exponentially small energy gaps~\cite{Albash2018,Nishimori2017,Hormozi2017,Susa2018}.
They accelerate an adiabatic process and contribute to better performance of computation.
Constrained quantum annealing (CQA) is a different approach to provide good performance~\cite{Hen2016a,Hen2016b,Kudo2018}.
In the CQA approach, the driver Hamiltonian is chosen so that hard constraints are naturally satisfied.
Practical optimization problems often have hard constraints.
To impose constraints, in a standard approach, 
penalty terms are added to the problem Hamiltonian.
The CQA approach does not require such extra terms.
The approach is also beneficial because it restricts the Hilbert space to a subspace with a considerably small dimension.
The dimension reduction enables us to perform real-time quantum simulations~\cite{Kudo2018}.

% Anderson localization
Localization is a potential factor to cause inefficiency of QA.
For example, localization may obstruct attaining possible solutions and lower the performance. 
Altshuler \textit{et.~al.} pointed out that a phenomenon similar to Anderson localization makes adiabatic quantum optimization fail~\cite{Altshuler2010}.
The Hamiltonian of QA with $N$ spins can be regarded as the system of a single quantum particle that moves between the vertices of an $N$-dimensional hypercube.
Each spin configuration $\{\sigma_i^z\}$ corresponds to a vertex of the hypercube.
The problem and driver Hamiltonians correspond to disordered potential and hopping terms, respectively.
Hence, the total Hamiltonian of QA describes a sort of the well-known Anderson model.
While Anderson localization can occur in a QA process,
the CQA of graph coloring provides another viewpoint of localization.
In the CQA of graph coloring, the total Hamiltonian is analogous to an ensemble of tight-binding chains under disordered fields~\cite{Kudo2018}.
The system is considered to be almost independent spin chains interacting weakly with other chains in the early stages of CQA.
In the late stage, as inter-chain interaction increases with time,
spins are affected by effective fields arising from neighboring chains.

% Aim of this paper
In this paper, we investigate localization phenomena in the CQA of graph coloring.
We focus on characteristics of localization in the CQA, instead of localization transition.
The localization transition in QA is a potential phase transition that is related to computational complexity.
Several works suggested that computational complexity should be related to phase transitions in QA~\cite{Amin2009,Young2010,Takahashi2019}.
However, it is not straightforward to find the order parameter for detecting such a phase transition, and discussion is often based on model-specific results.
Moreover, we cannot assume a close connection between localization transition and computational complexity in the CQA of graph coloring.
While localization is expected to occur in the middle of a QA process, the energy gap, which is related to computational efficiency, decreases monotonically in the CQA process~\cite{Kudo2018}.

% MBL and entanglement
This paper aims to provide a new viewpoint and a suitable method to analyze localization phenomena in the CQA of graph coloring.
Our study is based on numerical simulations of effective fields and entanglement in small quantum systems. 
Entanglement is often employed to characterize the many-body localization (MBL) transition~\cite{Znidaric2008,Bardarson2012,Serbyn2013a,Serbyn2013b,Bera2015,Bera2016,Enss2017,Kjall2014,Luitz2015,Khemani2017a,Khemani2017b,Zhang2018}.
MBL is localization in a quantum many-body system, and one can regard it as a sort of Anderson localization in Fock space~\cite{Welsh2018,Logan2019,Ghosh2019,Roy2019a,Roy2019b}.
In the MBL phase, where disorder is relatively strong compared with many-body interaction, quantum dynamics is nonergodic and breaks the eigenstate thermalization hypothesis (ETH)~\cite{Deutsch1991,Srednicki1994,Rigol2008}.
Although it is unclear whether MBL occurs in QA, entanglement is a useful quantity to characterize localization.
In this paper, we employ concurrence as an entanglement measure.
Concurrence, which is often used for measuring entanglement in a mixed state, is a quantity to measure pairwise entanglement between two spin-$1/2$ spins~\cite{Hill1997,Amico2008,Bera2016}.
This measure is useful for discussion from the viewpoint of localization in a spin chain under effective disordered fields.
Plotting of concurrence as a function of the relative strength of disorder, we clarify the difference of characteristics between the entanglement in the CQA and that in a disordered tight-binding chain.

% Outline of this paper
The remainder of the paper is organized as follows.
In Sec.~\ref{sec:model}, the model and numerical methods of the CQA of graph coloring are outlined, 
and effective fields and concurrence are defined.
The intra-chain concurrence and the concurrence in disordered chains are demonstrated in Secs.~\ref{sec:cncr} and \ref{sec:cncr_ds}, respectively.
The comparison of their results reveals notable features of localization in the CQA of graph coloring.
Conclusions are given in Sec.~\ref{sec:conc}.

\section{Models and methods}
\label{sec:model}

\subsection{Constrained quantum annealing}

We focus here on the CQA of graph coloring.
Graph coloring consists of coloring the nodes of a graph such that nodes directly connected through an edge do not share the same color.
When coloring a graph $G=(V,E)$ with $q$ available colors,
the classical Hamiltonian is given by
\begin{equation}
 H_{\rm cl} =\sum_{(ij)\in E}\sum_{a=1}^q
\frac{S_{i,a}+1}{2}\frac{S_{j,a}+1}{2},
\label{eq:H_cl}
\end{equation}
where $i$ and $j$ represent nodes $V=\{1,\ldots,N\}$, 
and $(ij)\in E$ denotes the edge connecting the pair of nodes $i,j\in V$.
If node $i$ is colored $a$, $S_{i,a}=1$; otherwise, $S_{i,a}=-1$.
Thus, Eq.~\eqref{eq:H_cl} counts the number of edges that connect nodes with the same color.
The quantum version of Eq.~\eqref{eq:H_cl} is the problem Hamiltonian,
given by
\begin{equation}
 H_{\rm p}=J\sum_{(ij)\in E}\sum_{a=1}^q \sigma^z_{i,a}\sigma^z_{j,a},
\label{eq:H_p}
\end{equation}
where $\sigma^z_{i,a}$ denotes the Pauli matrix of the component $z$, 
and $J$ has a unit of energy ($J=1$ in the simulations below).
As each node can only have one color, the required constraint is $\sum_{a=1}^q\sigma^z_{i,a}=2-q$.
In standard QA approaches,
the penalty term to satisfy this constraint is often incorporated into the problem Hamiltonian~\cite{Gatian2012,Bian2013,Gaitan2014,Lucas2014,Rieffel2015}.
In the CQA approach, however, 
we choose the driver Hamiltonian so that the constraint is satisfied consistently, instead of adding a penalty term.
Here, we give the driver Hamiltonian for this problem as
\begin{equation}
 H_{\rm d}=-J\sum_{i=1}^N\sum_{a,b}
\left(\sigma^x_{i,a}\sigma^x_{i,b}+\sigma^y_{i,a}\sigma^y_{i,b}\right).
\label{eq:H_d}
\end{equation}
In the following, we consider two types of summations over $a$ and $b$: 
nearest-neighbor (NN) and fully-connected (FC) types.
In the NN type, the summation over $a,b$ in the NN type is limited to $b=a+1$, and periodic boundary conditions are imposed. 
The summation in the FC type is taken for all combinations of $a<b$.
The total Hamiltonian is given as
\begin{align}
 H(s) = sH_{\rm p} + (1-s)H_{\rm d},
\end{align}
where $s$ is a time-dependent parameter.
The QA process starts at $s=0$ and ends at $s=1$.
As the total Hamiltonian is a type of $XXZ$ model, 
the magnetization $\sum_{a=1}^q\sigma^z_{i,a}$ is conserved for each $i$.

% schematics of the model
% --- Fig: model ---
\begin{figure}[tb]
\centering 
\includegraphics[width=8cm,clip]{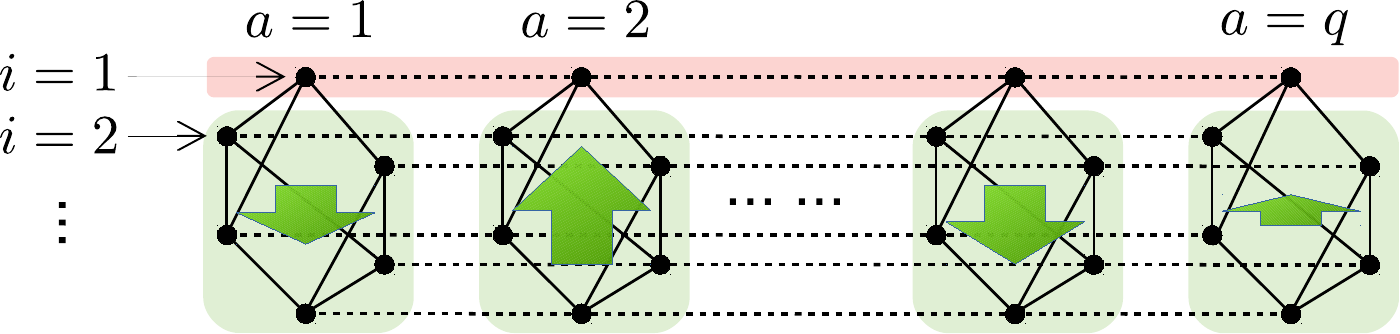}%
\caption{\label{fig:chain}
Schematic of the model. Thick arrows represent effective local fields.
The solid and dashed lines correspond to the problem Hamiltonian \eqref{eq:H_p} and the driver Hamiltonian \eqref{eq:H_d}, respectively.
If we focus on the solid line (spin chain) of $i=1$, the nodes neighboring the chain play the role of effective fields.   
}
\end{figure}

The model is a kind of an ensemble of tight-binding chains.
In Fig.~\ref{fig:chain}, which is a schematic of the model,
each dashed line corresponds to a tight-binding chain.
Because each chain only contains one up-spin, there is no interaction between up-spins in a chain.
As represented in Fig.~\ref{fig:chain},
each site of a chain (a dashed line) interacts with the corresponding site of neighboring chains through the problem Hamiltonian (solid lines).
We can consider that the tight-binding chain is affected by effective local fields (represented by arrows) that arise from the spins in its neighboring chains.

% implementation: annealing schedule
In the CQA approach, we only use the subspace of $\sum_{a=1}^q\sigma^z_{i,a}=2-q$, which satisfies the constraint.
The initial state is the lowest-energy state of $H_{\rm d}$ in the subspace.
Although this initial state is not the global ground state, the state can be prepared in the whole Hilbert space of the system by adding additional Zeeman term with an appropriate magnetic field~\cite{Hen2016b}.
In the following sections, we refer to the lowest energy state in the subspace as the ground state.
Since the time evolution in a CQA process is ideally supposed to be adiabatic, we deal with the instantaneous ground state, in this paper, instead of real-time Schr\"{o}dinger evolution.
We obtain the ground state using the Lanczos method.

% advantage of the CQA: time evolution
The benefits of the CQA approach are not only that constraints are naturally satisfied without penalty terms but also
that the dimension of the Hilbert space reduces considerably.
The dimension reduction is essential for obtaining the instantaneous ground state.
In a standard approach, one needs $2^{qN}$ dimensions for coloring a graph of $N$ nodes with $q$ colors.
However, with the CQA approach described above, the number of required dimensions is reduced to only $q^N$.

% regular random graphs
The graphs considered here are regular random graphs.
Each node has the same degree in a regular graph.
Regular random graphs are selected randomly from regular graphs.
In this paper, we mainly consider four-coloring ($q=4$) of small regular graphs ($N=8$) of degree $c=2,3,4$.
Those graphs are expected to be colorable, 
since the transition threshold below which regular graphs are colorable in the large-$N$ limit is $c_s=10$ (for $q=4$)~\cite{Krzakala2004,Zdeborova2007}.

\subsection{Effective fields}

When $s>0$, spins are affected by effective fields arising from the interaction between neighboring nodes in a given graph.
Note that the problem Hamiltonian~\eqref{eq:H_p} can be rewritten as
\begin{align}
 H_{\rm p} = \sum_{i=1}^N\sum_{a=1}^q
\left( \sum_{j=1}^N J_{ij}\sigma_{j,a}^z \right)\sigma_{i,a}^z,
\end{align}
where $J_{ij}=J/2$ for $(ij)\in E$, otherwise $0$.
The average of effective fields, which is time dependent, is defined as
\begin{align}
 \langle \hat{h}^{\rm eff}_{i,a}(s) \rangle
\equiv \langle\Psi(s)| \hat{h}^{\rm eff}_{i,a}(s) |\Psi(s)\rangle,
\end{align}
where $|\Psi(s)\rangle$ is the wavefunction at $s$ and 
\begin{align}
 \hat{h}^{\rm eff}_{i,a}(s) = s\sum_{j=1}^N J_{ij}\sigma_{j,a}^z.
\label{eq.h_eff.0}
\end{align}
Similarly, the fluctuation of effective fields is defined as 
\begin{align}
 \Delta^{\rm eff}(s) \equiv \sqrt{ \langle(\hat{h}^{\rm eff}_{i,a}(s))^2\rangle
- \langle \hat{h}^{\rm eff}_{i,a}(s) \rangle^2 }.
\end{align}
Considering regular random graphs of degree $c$, 
the average of effective fields at $s=1$ is estimated as
\begin{align}
 \langle \hat{h}^{\rm eff}_{i,a} \rangle
=\frac{cJ}{2}\left(\frac{1}{q}-\frac{q-1}{q}\right)
=-\frac{cJ(q-2)}{2q}.
\label{eq:h_eff}
\end{align}
Similarly, we have
$\langle (\hat{h}^{\rm eff}_{i,a})^2 \rangle=(cJ^2/4)[1+(c-1)(q-2)^2/q^2]$.
Thus, the estimated value of the fluctuation at $s=1$ is written as
\begin{align}
 \Delta_1 = \frac{J}{2}\sqrt{c\left[1-\left(q-2\right)^2/q^2\right]}.
\label{eq:D_eff}
\end{align}
If the population at each site ($a =1, \ldots, q$) 
in each chain $(i=1, \ldots, N)$ is equal, 
the average of effective fields is equal to Eq.~\eqref{eq:h_eff} multiplied by $s$.
In the same situation, the fluctuation of effective fields is expected to increase as $\Delta^{\rm eff}(s)=s\Delta_1$.
In the numerical simulations below, we take $i=a=1$. 

% --- Fig: graph (N=6) ---
\begin{figure}[tb]
 \centering
 \includegraphics[width=6cm,clip]{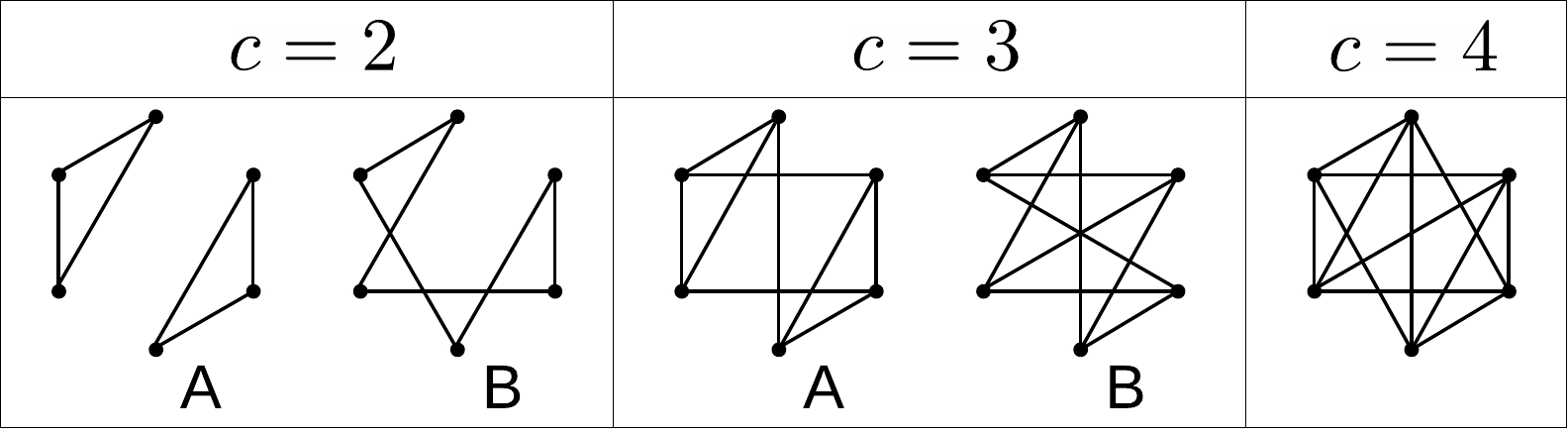}%
\caption{\label{fig:graphs_n6}
Selected graphs of $N=6$ and degree $c=2,3,4$.
}
\end{figure}

% --- Fig: effective field ---
\begin{figure}[tb]
\centering
 \includegraphics[width=5cm,clip]{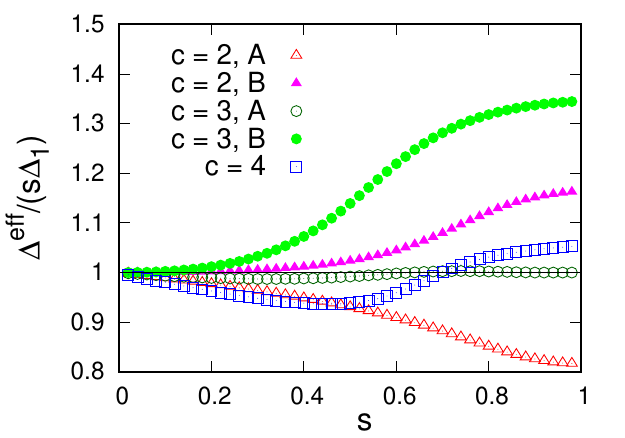}%
\caption{\label{fig:field}
The fluctuation of effective fields $\Delta^{\rm eff}$ divided by $s\Delta_1$ as a function of $s$ for the graphs in Fig.~\ref{fig:graphs_n6}.
Here, $\Delta_1$ is given by Eq.~\eqref{eq:D_eff} with $J=1$ and $q=4$.
}
\end{figure}

% heff for n=6
We confirmed in numerical simulations that the average of effective fields is proportional to Eq.~\eqref{eq:h_eff} as expected.
However, the fluctuation of effective fields $\Delta^{\rm eff}$ shows nonlinear growth.
For selected graphs of $N=6$, which are shown in Fig.~\ref{fig:graphs_n6},
we demonstrate $\Delta^{\rm eff}$ divided by $s\Delta_1$ as a function of $s$ in Fig.~\ref{fig:field}.
If $\Delta^{\rm eff}$ increases as expected, $\Delta^{\rm eff}/(s\Delta_1)$ should be unity.
The nonlinear growth in $\Delta^{\rm eff}$ implies that the population in each site is not equal.
A large deviation from $\Delta^{\rm eff}/(s\Delta_1)=1$ indicates localization.

% relative strength of disorder
Considering the analogy between the model and a disordered tight-binding chain, we can regard $\Delta^{\rm eff}(s)$ as disorder strength. 
Since $\Delta^{\rm eff}(s)$ is not proportional to $s$, 
the dependence on $s$ is not suitable for the discussion of localization from the viewpoint of the analogy to a disordered chain.
Instead, we consider the dependence on $\Delta^{\rm eff}/(1-s)$, 
which corresponds to the relative strength of disorder, 
where $\Delta^{\rm eff}(s)$ and $1-s$ represent disorder and hopping strengths, respectively. 

\subsection{Concurrence}

We here employ concurrence as a measure of pairwise entanglement, although half-chain entanglement entropy is used in most studies on one-dimensional spin systems.
Since concurrence is a quantity defined between two spins, it is independent of the dimension and structure of a system.
Therefore, concurrence is useful for comparison between the entanglement in the CQA model and that in a disordered chain.

The concurrence $C_{i,j}$ in spins $i$ and $j$ is defined from the eigenvalues of the matrix $\rho_{ij}\tilde{\rho}_{ij}$,
where $\rho_{ij}$ is the reduced density matrix, and 
$\tilde{\rho}_{ij}=\sigma^y\otimes\sigma^y\rho_{ij}^*\sigma^y\otimes\sigma^y$.
Note that the wavefunction $|\Psi(s)\rangle$ is a pure state, although $\rho_{ij}$ is a mixed state in general.
Suppose that the eigenvalues are 
$\lambda_1\ge\lambda_2\ge\lambda_3\ge\lambda_4$,
then the concurrence is $C_{i,j}=\max
(\sqrt{\lambda_1}-\sqrt{\lambda_2}-\sqrt{\lambda_3}-\sqrt{\lambda_4},\; 0)$.
From the conservation of magnetization, 
the concurrence in the systems considered here can be expressed in a simple form~\cite{OConnor2001,Bera2016}:
\begin{equation}
 C_{i,j} = 2\max(|z|-\sqrt{xy},\; 0),
\end{equation}
where $z=\langle\uparrow\downarrow|\rho_{ij}|\downarrow\uparrow\rangle$,
$x=\langle\uparrow\uparrow|\rho_{ij}|\uparrow\uparrow\rangle$, and
$y=\langle\downarrow\downarrow|\rho_{ij}|\downarrow\downarrow\rangle$.

\section{Intra-chain concurrence}
\label{sec:cncr}

In our system, a spin has two indices, i.e., node $i$ and color $a$.
Let $C(i,a;\; j,b)$ denote the concurrence in spins $(i,a)$ and $(j,b)$.
We define the intra-chain concurrence of the $i$th chain as
\begin{equation}
 C_i^{\rm ch} = \frac12 \sum_{a=1}^q C(i,a;\; i,a+1),
\label{eq:C_ch}
\end{equation}
which is scaled so that $C_i^{\rm ch}=1$ at $s=0$.
Equation~\eqref{eq:C_ch} describes the nearest-neighbor concurrence in the $i$th chain.

% --- Fig: intra-chain concurrence (linear plot) ---
\begin{figure}[tb]
\centering
 \includegraphics[width=4cm,clip]{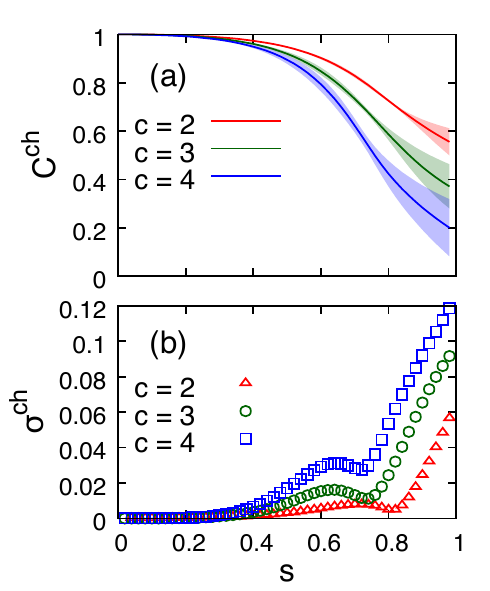}%
\caption{\label{fig:C_lin}
The average $C^{\rm ch}$ (a) and standard deviation $\sigma^{\rm ch}$ (b) of the intra-chain concurrence plotted as functions of $s$.
The driver Hamiltonian is the NN type.
The error bars in (a) represent $\sigma^{\rm ch}$.
The number of nodes is $N=8$, and $c$ is the degree of a regular graph.
}
\end{figure}

% --- Fig: intra-chain concurrence (log-log plot for NN and FC)---
\begin{figure}[tb]
\centering
 \includegraphics[width=8cm,clip]{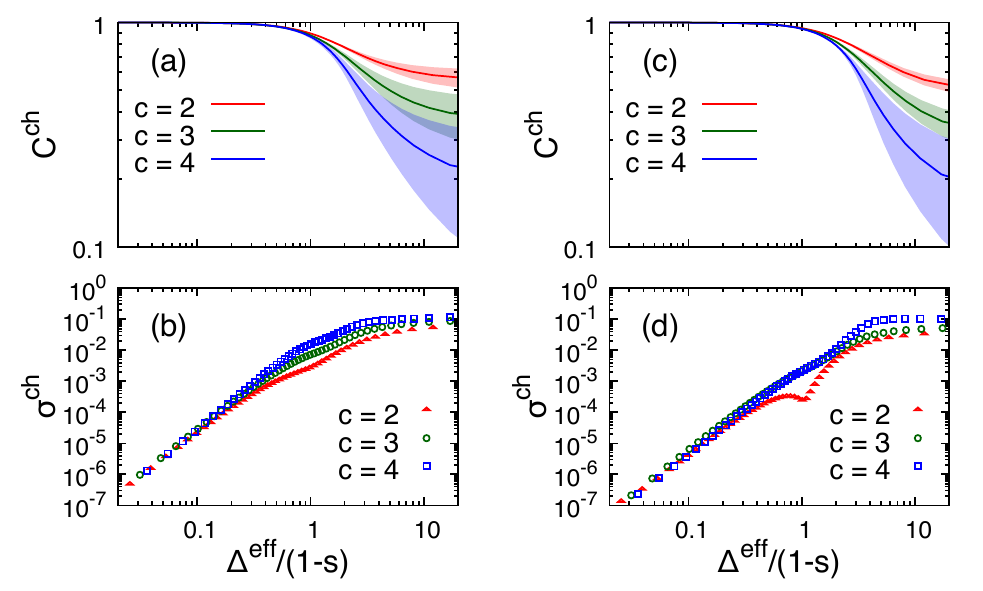}%
\caption{\label{fig:C_log}
The average $C^{\rm ch}$ and standard deviation $\sigma^{\rm ch}$ of the intra-chain concurrence are plotted in a log-log scale as functions of the relative strength of disorder $\Delta^{\rm eff}/(1-s)$.
The driver Hamiltonian is given by the NN type in (a)--(b) and the FC type in (c)--(d).
The error bars in (a) and (c) represent $\sigma^{\rm ch}$.
The numbers of nodes and colors are $N=8$ and $q=4$, respectively.
$c$ is the degree of a regular graph.
}
\end{figure}

Figure~\ref{fig:C_lin} plots the average $C^{\rm ch}$ and standard deviation $\sigma^{\rm ch}$ of the intra-chain concurrence for $N=8$ as functions of $s$.
The intra-chain concurrence, which is defined by Eq.~\eqref{eq:C_ch} with $i=1$ and $q=4$, is calculated using the instantaneous ground state for $1000$ random regular graphs with degree $c$. 
Here, the driver Hamiltonian is the NN type.
Since only one spin is up and the others are down in each chain, 
the concurrence in the ground state at $s=0$ (and $\Delta^{\rm eff}=0$) is maximal and $C(i,a;\; i,a+1)=2/q$, which leads to $C_i^{\rm ch}=1$.
While the average $C^{\rm ch}$ decreases monotonically,
the standard deviation $\sigma^{\rm ch}$ shows a nonmonotonic increase and has a trough after a small peak.
The trough corresponds to the point where many curves of intra-chain concurrence cross for the same-$c$ graphs.
The driver Hamiltonian $H_{\rm d}$ is dominant before the trough, and the problem Hamiltonian $H_{\rm p}$ becomes dominant after the trough.
In this sense, the trough is considered as a signature of the crossover from quantum to classical behavior.

Although Fig.~\ref{fig:C_lin} shows suggestive results, it is not suitable for the discussion from the viewpoint of the analogy to a disordered chain.
In Fig.~\ref{fig:C_log}, we plot the average $C^{\rm ch}$ and standard deviation $\sigma^{\rm ch}$ of the intra-chain concurrence as functions of the relative strength of disorder $\Delta^{\rm eff}/(1-s)$ in a logarithmic scale.
The driver Hamiltonian is the NN type in Figs.~\ref{fig:C_log}(a)--(b) and the FC type in Figs.~\ref{fig:C_log}(c)--(d).
In contrast to Fig.~\ref{fig:C_lin}(b), no trough appears in Fig.~\ref{fig:C_log}(b).
Instead, in Fig.~\ref{fig:C_log}(b), slow slopes appear around $\Delta^{\rm eff}/(1-s)\sim 1$ for different $c$.
The slow-slope region corresponds to the trough in Fig.~\ref{fig:C_lin}(b).
The regions before and after the slow slope are in the hopping-dominant ($\Delta^{\rm eff}<1-s$) and disorder-dominant ($\Delta^{\rm eff}>1-s$) regimes, respectively.
Similar behavior appears in the case of FC-type driver Hamiltonian, as shown in Figs.~(c)--(d).

% --- Fig: intra-chain concurrence (larger systems) ---
\begin{figure}[tb]
\centering
 \includegraphics[width=8cm,clip]{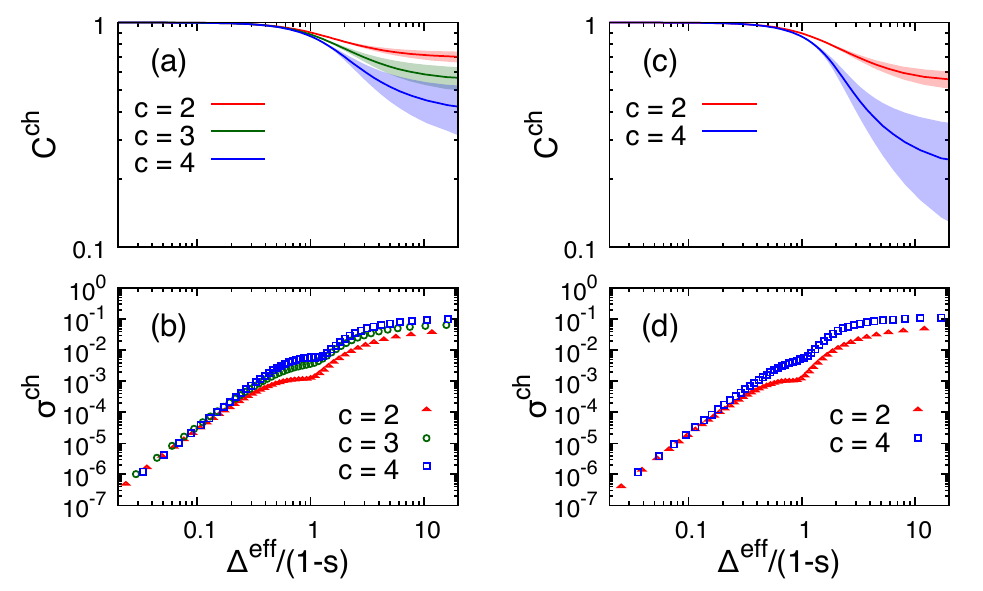}%
\caption{\label{fig:C_8594}
The average $C^{\rm ch}$ and standard deviation $\sigma^{\rm ch}$ of
the intra-chain concurrence are plotted in a log-log scale as functions of $\Delta^{\rm eff}/(1-s)$. 
The number of nodes and colors are, respectively, $N=8$ and $q=5$ in (a)--(b), and $N=9$ and $q=4$ in (c)--(d).
The error bars in (a) and (c) represent $\sigma^{\rm ch}$.
The driver Hamiltonian is the NN type. 
Note that the degree $c$ of a regular graph with an odd number of nodes is limited to even numbers.
}
\end{figure}

Dependence on the system size is twofold: the number of colors $q$ and that of nodes $N$.
When $q$ becomes large, the average $C^{\rm ch}$ of the intra-chain concurrence tends to maintain high values.
This behavior is demonstrated in Fig.~\ref{fig:C_8594}(a), where $N=8$ and $q=5$.
Since the occupation probability at each site is small for a large $q$, the concurrence between the neighboring sites is also small.
Since concurrence $C(i,a;\; i,a+1)$ of each spin pair is small, the difference in concurrence between different spin pairs is also small.
Then, $C(i,a;\; i,a+1)\simeq 2/q$ for all $a$, which implies $C_i^{\rm ch}\simeq 1$.
In contrast to the $q$ dependence, the dependence on $N$ is not remarkable.
The behavior of the intra-chain concurrence in Fig.~\ref{fig:C_8594}(c), which is for $N=9$ and $q=4$, looks similar to that of Fig.~\ref{fig:C_log}(a), which is for $N=8$ and $q=4$. 
The standard deviation $\sigma^{\rm ch}$ also shows similar behavior to Fig.~\ref{fig:C_log}(b), although slow-slop regions of $\sigma^{\rm ch}$ in Figs.~\ref{fig:C_8594}(b) and (d) are clearer than that in Fig.~\ref{fig:C_log}(b).

\section{Concurrence in disordered chains}
\label{sec:cncr_ds}

% Compare with the case of disordered spin chains
Next, we compare the results with those of disordered chains.
We give the Hamiltonian of a disordered chain as
\begin{align}
 H(s) = -2(1-s)J\sum_{a=1}^q\left(
c_a^\dagger c_{a+1} + c_{a+1}^\dagger c_a \right)
+ 2sJ\sum_{a=1}^q h_a c_a^\dagger c_a,
\end{align}
where $c_a^\dagger$ and $c_a$ are the creation and annihilation operators of spinless fermions, respectively, and $h_a$ denotes a local potential. 
Periodic boundary conditions are imposed.
There is only one particle in the chain, which corresponds to the one-up-spin condition.
We define the corresponding concurrence as
\begin{align}
 C^{\rm ds}=\frac12\sum_{a=1}^q C_{a,a+1},
\label{eq:C_ds}
\end{align}
which is scaled so that $C^{\rm ds}=1$ at $s=0$.

% --- Fig: concurrence of disorderd chains (uniformly random) ---
\begin{figure}[tb]
\begin{center}
 \includegraphics[width=8cm,clip]{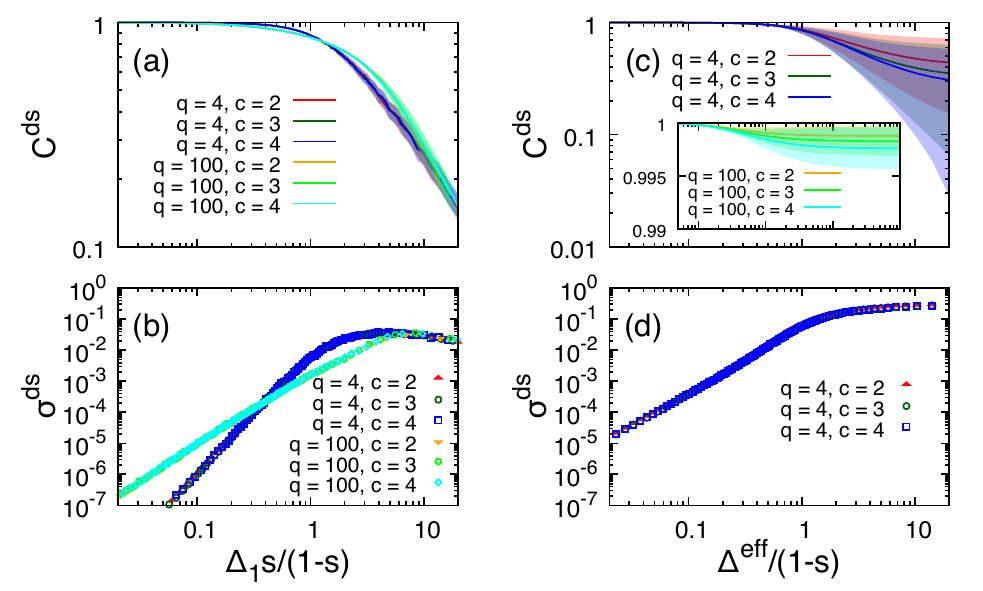}%
\caption{\label{fig:C_ds}
The average $C^{\rm ds}$ and standard deviation $\sigma^{\rm ds}$ of the concurrence of disordered chains defined by Eq.~\eqref{eq:C_ds} with $q=4$ and $100$.
The error bars in (a) and (c) represent $\sigma^{\rm ds}$.
In (d), $\sigma^{\rm ds}$ for $q=100$ is not plotted because $\sigma^{\rm ds}\simeq 0$.
(a)--(b) The local potential is composed of a random number drawn from a uniform distribution, 
and the curves are plotted as functions of $\Delta_1 s/(1-s)$, where $\Delta_1$ is given by Eq.~\eqref{eq:D_eff}.
(c)--(d) The local potential is given by Eq.~\eqref{eq:ha}, and the curves are plotted as functions of $\Delta^{\rm eff}/(1-s)$.
}
\end{center}
\end{figure}

The conventional manner to give a local random potential is to draw a random number from a uniform distribution $[\mu-w, \mu+w]$.
Since the average of effective fields is given by Eq.~\eqref{eq:h_eff} with $J=1$, we take $\mu = -c(1/2-1/q)s$.
Adjusting the standard deviation of random potentials to the fluctuation of effective fields, we take $w=\sqrt{3}\Delta_1s$, where $\Delta_1$ is given by Eq.~\eqref{eq:D_eff}.
Figures~\ref{fig:C_ds}(a)--(b) demonstrate the average $C^{\rm ds}$ and standard deviation $\sigma^{\rm ds}$ of the concurrence for $q=4$ and $q=100$.
The average is taken over 1000 samples for each point.
In contrast to the intra-chain concurrence,
neither the average nor standard deviation shows the dependence on $c$, and the standard deviation is quite small.
While the average $C^{\rm ds}$ exhibits power-law decay in the strong-disorder regime, the standard deviation $\sigma^{\rm ds}$ does a power-law increase in the weak-disorder regime.
The slight decay in $\sigma^{\rm ds}$ in the strong-disorder regime comes from the fact that the average $C^{\rm ds}$ is substantially small and is nonnegative.
Although the decay in $\sigma^{\rm ds}$ starts earlier for $q=100$ than for $q=4$, the dependence on $q$ is not significant. 

Here, a question arises:
What causes the difference between the intra-chain concurrence and the concurrence in a disordered chain?
The difference in the characteristics of disorder is a possible cause.
Here, we change the method to give a local random potential so that $sh_a$ mimics an effective field, Eq.~\eqref{eq.h_eff.0}.
The local potential is now given by a discrete random number,
\begin{align}
 h_a = \frac12\sum_{i=1}^c m_i,
\label{eq:ha}
\end{align}
where $m_i=1$ with a probability $1/q$, and $m_i=-1$ with a probability $1-1/q$.
Figures~\ref{fig:C_ds}(c)--(d) demonstrate the average $C^{\rm ds}$ and standard deviation $\sigma^{\rm ds}$ of the concurrence.
The average is taken over 1000 chains, namely, 
1000 sets of $h_a$ ($a=1,\ldots, q$). 
The standard deviation of each set of random potentials $h_a$ (multiplied by $s$) gives $\Delta_{\rm eff}$.

The results shown in Figs.~\ref{fig:C_ds}(c)--(d) have several characteristics similar to those of the intra-chain concurrence.
First, the dependence on $q$ is significant in Figs.~\ref{fig:C_ds}(c)--(d), compared to Figs.~\ref{fig:C_ds}(a)--(b).
The behavior that $C^{\rm ds}\simeq 1$ for a large $q$ is the same as that of the intra-chain concurrence.
Second, we see the $c$-dependence of the average $C^{\rm ds}$ of the concurrence in a disordered chain in Fig~\ref{fig:C_ds}(c), although it is weaker than that of the intra-chain concurrence.
This result implies that the discrete randomness or discrete nature of effective fields is a possible reason behind the $c$-dependence of the average of the concurrence.

In contrast, the standard deviation $\sigma^{\rm ds}$ of the concurrence in Fig.~\ref{fig:C_ds}(d) is independent of $c$. 
Moreover, the curve of $\sigma^{\rm ds}$ has no slow-slope region around $\Delta_{\rm eff}/(1-s)\sim 1$ and only shows a slow increase in the strong-disorder regime.
In the model of the CQA of graph coloring, spins in neighboring chains can interact through the problem Hamiltonian.
In a disordered chain, however, there is no particle-particle interaction.
Thus, the inter-chain interaction in the CQA model is a possible reason for the difference between the characteristics of the standard deviation of the intra-chain concurrence and that of the concurrence in a disordered chain.

\section{Conclusions}
\label{sec:conc}

We investigated localization phenomena in the CQA of graph coloring from the viewpoint of analogy to a tight-binding chain under effective fields.
Effective fields that arise from neighboring chains behave as disorder and cause localization during a CQA process.
The fluctuation of effective fields, i.e., disorder strength, is not directly proportional to the annealing parameter $s$.
We analyzed the intra-chain concurrence as a function of relative disorder strength instead of $s$ and found a remarkable feature:
The standard deviation of the intra-chain concurrence exhibits a slow-slope region around the point where the disorder strength balances with that of hopping. 
This feature does not appear in the concurrence for corresponding disordered chains, which implies that inter-chain interaction is a possible cause of the slow-slope region.
The method focusing on the dependence on the relative disorder strength has the potential to reveal other characteristics that have not been captured in the dependence of $s$.
Even though the above simulations are for small-size systems, those results based on the diagonalization of a quantum Hamiltonian make some contribution to understanding localization phenomena in QA.
Investigation in much larger systems is highly desirable for better understanding.

\begin{acknowledgments}
The author would like to thank Kazutaka Takahashi for useful discussions.
This work is partially supported by JSPS KAKENHI Grant Number JP18K11333, and the research grant from the Inamori Foundation.
\end{acknowledgments}

% Create the reference section using BibTeX:
%\bibliography{qma.bib}

\end{document}